\documentclass[aps,prl,reprint,superscriptaddress]{revtex4-1}
\pdfoutput = 1
\usepackage{filecontents}
\usepackage{hyperref} 
\usepackage{tikz} 
\usepackage{amsmath} 
\usepackage{verbatim} 
\usepackage{braket}
\begin{document}

\title{A neutron scattering measurement of crystalline-electric fields in magnesium rare-earth selenide spinels}

\author{D. Reig-i-Plessis}
\affiliation{Department of Physics and Materials Research Laboratory, University of Illinois at Urbana-Champaign, Urbana, Illinois, 61801, USA}
\author{A. Cote}
\affiliation{Department of Physics and Materials Research Laboratory, University of Illinois at Urbana-Champaign, Urbana, Illinois, 61801, USA}
\author{S. van Geldern}
\affiliation{Department of Physics and Materials Research Laboratory, University of Illinois at Urbana-Champaign, Urbana, Illinois, 61801, USA}
\author{R. D. Mayrhofer}
\affiliation{University of Rochester, Rochester, New York, 14627, USA}
\author{A. A. Aczel}
\affiliation{Neutron Scattering Division, Oak Ridge National Laboratory, Oak Ridge, Tennessee, 37831, USA}
\affiliation{Department of Physics and Astronomy, University of Tennessee, Knoxville, Tennessee, 37996, USA}
\author{G. J. MacDougall}
\email{gmacdoug@illinois.edu}
\affiliation{Department of Physics and Materials Research Laboratory, University of Illinois at Urbana-Champaign, Urbana, Illinois, 61801, USA}

\begin{abstract}

The symmetry of local moments plays a defining role in the nature of exotic ground states stabilized in frustrated magnetic materials. We present inelastic neutron scattering (INS) measurements of the crystal electric field (CEF) excitations in the family of compounds MgRE$_2$Se$_4$ (RE $\in$ $\{$Ho, Tm, Er and Yb$\}$). These compounds form in the spinel structure, with the rare earth ions comprising a highly frustrated pyrochlore sublattice. Within the symmetry constraints of this lattice, we fit both the energies and intensities of observed transitions in the INS spectra to determine the most likely CEF Hamiltonian for each material and comment on the ground state wavefunctions in the local electron picture. In this way, we experimentally confirm MgTm$_2$Se$_4$ has a non-magnetic ground state, and MgYb$_2$Se$_4$ has effective $S=\frac{1}{2}$ spins with $g_\parallel = 5.188(79)$ and $g_\perp = 0.923(85)$. The spectrum of MgHo$_2$Se$_4$ indicates a ground state doublet containing Ising spins with $g_\parallel = 2.72(46)$, though low-lying CEF levels are also seen at thermally accessible energies $\delta_E = 0.591(36)$, 0.945(30) and 2.88(7)~meV, which can complicate interpretation. These results are used to comment on measured magnetization data of all compounds, and are compared to published results on the material MgEr$_2$Se$_4$.
\end{abstract}

\maketitle

\section{Introduction}

The strategic combination of frustrated lattice geometries and strong local-ion anisotropy is a well-established route for stabilizing novel spin states in quantum materials\cite{01_moessner_review_frustration,greedan2006frustrated,12_castelnovo_review_spin_ice}.
This fact endows a special significance to the local crystal electric field (CEF) Hamiltonian of magnetic moments\cite{12_Bertin_CEF_across_R2Ti2O7}, which dictates size, dimensionality, and allowed interactions in effective spin systems at low temperatures. In $f$-electron systems in particular, spin and orbital degrees of freedom are strongly coupled, leading to dramatic changes in the CEF splittings depending on the number of valence electrons.
Additionally the small radius and shielding of the $f$-electron orbitals leads to small CEF splittings, necessitating full consideration of excited states. As a result, even within a family of closely related structures, each change in the number of valence electrons creates an entirely new effective spin system and leads to a wide range of interesting behaviors.

A particularly important example from recent years is the so-called `227' family of pyrochlore oxides, A$_2$RE$_2$O$_7$ (A = cation, RE = rare earth), where rare-earth moments occupy a frustrated sublattice of corner-sharing tetrahedra \cite{10_gardner}.
The phase diagram of these materials is exceedingly rich, and includes such diverse states as non-collinear order\cite{01_champion_Gd2Ti2O7}, spin glass\cite{86_greedan_Y2Mo2O7}, classical spin liquid \cite{99_gardner, gardner2001neutrontb2ti2o7}, and both classical\cite{97_harris, ramirez1999zero, bramwell2001spin} and quantum\cite{applegate2012vindication, ross2011quantum, pan2014low, 16_sibille_PrHfO_QSL, 17_wen_QSL_disorderd_PrZrO, 17_savary_QSL_via_disorder} variants of the ``spin ice'' states.
This variety mirrors the number of different local symmetries selected through the interaction between the valence shell and the CEF\cite{12_Bertin_CEF_across_R2Ti2O7}, which spans possibilities from strongly Ising-like \cite{97_harris,bramwell2001spin} to XY \cite{17_Hallas_pyrochlore_review,18_hallas_XY_pyrochlores_review} to Heisenberg \cite{01_champion_Gd2Ti2O7,08_stewart_Gd2Sn2O7_dynamics,06_Wills_ordering_Gd2Sn2O7} moments. Virtual transitions associated with low-lying CEF states have further been credited with inducing quantum fluctuations\cite{07_Molavian_virtual_CEF_excitations,16_Rau_order_by_virtual_CEF}, while effects of multipolar local ion symmetries are suggested to lead to unexpected spin orders\cite{15_Lhotel_DO_ordering_Nd2Zr2O7}, quantum spin ice \cite{14_Huang_QSI_w_DO_orbitals,17_Li_dipole-octupoles_pyrochlore}, or other enriched spin liquid  states\cite{15_Sibille_QSL_candidate_Ce2Sn2O7}.

\begin{figure}[tbh]
\centering
\includegraphics[width=\columnwidth]{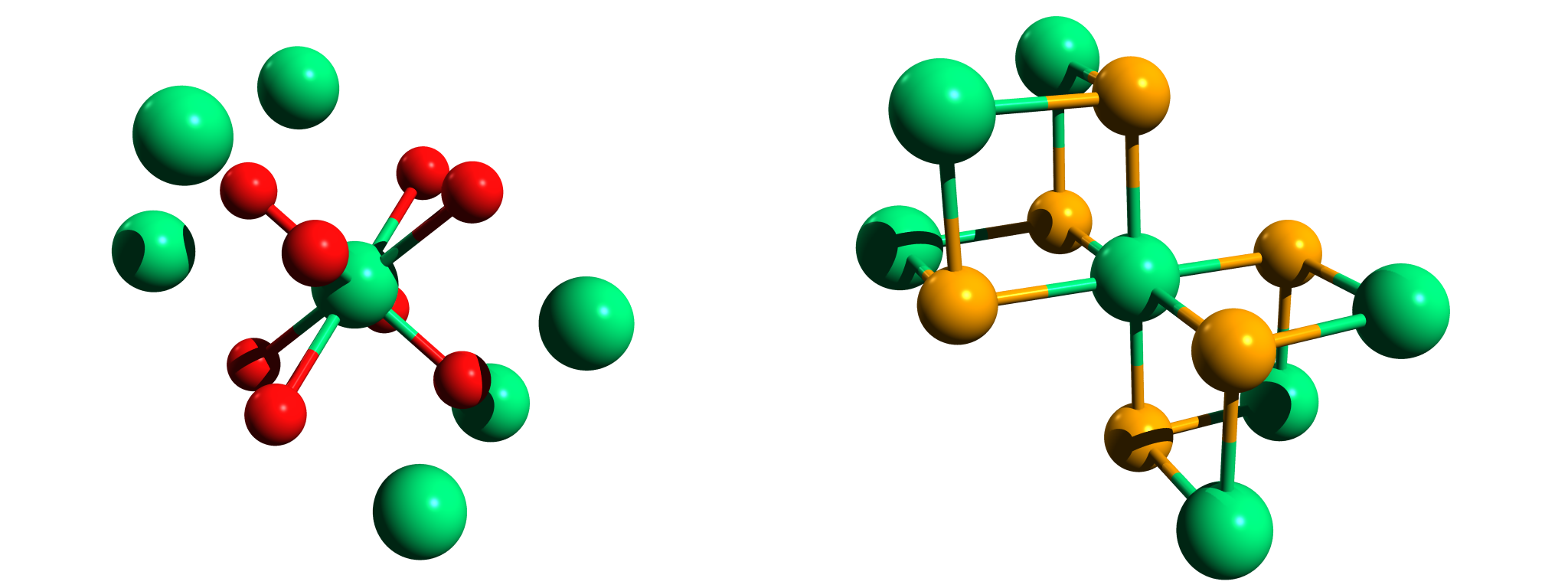}
\caption{Local environment of the A-site in the pyrochlore oxides (left) and the local environment of the RE-site in the chalcogenide spinels (right). The rare earth ion is displayed in green, O$^{2-}$ in red, and chalcogenide X$^{2-}$ in orange.}
\label{fig:local_env}
\end{figure}

This wide variety of exotic states in the single family of isostructural 227 oxides has generated strong interest in other materials in which rare earth moments comprise pyrochlore sublattices, with potentially new CEF environments.
Of these, perhaps most prominent have been the rare-earth (RE) spinel chalcogenides:  ARE$_2$X$_4$, with X $\in$ \{S, Se\}.
Both spinel and 227 pyrochlore families have global $Fd\bar{3}m$ symmetry and the magnetic cations comprise identical frustrated sublattices.
The local environment of the moments are substantially different, however, as demonstrated in Fig~\ref{fig:local_env}.
The A-site cations in 227 pyrochlores are surrounded by heavily-distorted cubes of O$^{2-}$ atoms, with a large trigonal distortion along the $\langle 111\rangle$ directions.
In contrast, the moments in spinels lie at the center of nearly perfect octahedral cages of X$^{2-}$ anions, with trigonal fields arising from both the compression or expansion of the REX$_6$ octahedra and the antiprism of neighboring cation sites\cite{mun2014, yaresko2008, reig2018spin}.
This substantial difference in local environment allows specific rare earth ions to adopt a drastically different symmetry in the two material families. For example, Er$^{3+}$ has XY-like moments in the `227' pyrochlores\cite{07_Poole_ordering_Er2Ti2O7} and Ising-like moments in the spinels\cite{reig2018spin}.

Among ternary rare-earth chalcogenides, the spinel phases have been confirmed for compounds with A $\in$ \{Cd, Mg\} and  RE $\in$ \{Ho, Er, Tm, Yb\}\cite{65_flahaut,66_flahaut_erratum}.
Earliest measurements  of material properties were performed in the 1960's-1980's, and  employed mostly X-ray diffraction (XRD) \cite{65_flahaut,66_flahaut_erratum,72_Fuji_Mag_CdRE2X4,80_Ben-Dor_CdRE2S_series}, magnetization\cite{72_Fuji_Mag_CdRE2X4,74_Pokrzywnicki_CEF_CdYb2Se4,75_Pokrzywnicki_CEF_CdYb2Se4,77_pokrzywnicki_mag_tm2se3_CdTm2Se4,80_Ben-Dor_CdRE2S_series,88_pawlak_Mag_CEF_Yb} and M\"ossbauer spectroscopy\cite{81_ben-dor_CdEr2Se4_mag_and_mossbauer}. X-ray diffraction measurements\cite{65_flahaut,66_flahaut_erratum,72_Fuji_Mag_CdRE2X4,80_Ben-Dor_CdRE2S_series} confirmed early on the ideal cubic $Fd\bar{3}m$ structure for the entire series, and further indicated that this high symmetry persists to the lowest measured temperatures. This observation stands in contrast to the symmetry-lowering cooperative structural transitions that are typically observed in spinel oxides\cite{10_lee_spinel_review,07_suzuki_OO_MnV2O4,08_garlea_OO_MnV2O4,99_yamada_JT_Mn}.

Original magnetization studies reported no order in the compound CdHo$_2$X$_4$ above 2~K \cite{72_Fuji_Mag_CdRE2X4}, and at least one argued for a spin singlet ground state on the basis of an observed temperature independent paramagnetic signal \cite{80_Ben-Dor_CdRE2S_series}.
Early work on CdEr$_2$X$_4$ claimed the onset of magnetic order in the temperature region $T$ = 4 - 10 K based on magnetization \cite{72_Fuji_Mag_CdRE2X4} and M\"ossbauer spectroscopy \cite{81_ben-dor_CdEr2Se4_mag_and_mossbauer}, though these reports stood in conflict with one another and failed to appreciate the consequences of local spin anisotropy on their data.
Early reports on CdYb$_2$Se$_4$ provided a more comprehensive analysis, and determined the CEF excitation energies of  20.6~meV and 40.7~meV using a model which accounted for an octahedral environment of Se-anions\cite{74_Pokrzywnicki_CEF_CdYb2Se4,75_Pokrzywnicki_CEF_CdYb2Se4}, though important contributions from neighboring cations were ignored.
The same study estimated a nearest-neighbor exchange energy to be around $J\approx$ 2.2~K. Studies of CdTm$_2$Se$_4$ concluded having a spin singlet ground state, consistent with expectations\cite{80_Ben-Dor_CdRE2S_series,77_pokrzywnicki_mag_tm2se3_CdTm2Se4}.

In recent years, interest in RE spinel chalcogenides has seen a revival, with a sharper focus on the frustrated nature of interactions\cite{05_Lau_RE_chalcogenide_spinels} and the resultant potential for novel forms of magnetism\cite{10_Lago_CdEr2Se4_heat_cap,15_Legros_mossbauer_CdEr2Se4,15_Yaouanc_CdHo2S4,17_Higo_AYb2S4,17_Reotier_order_and_tunneling_in_CdYb2Se4,18_Gao_CdEr2Se4_fennel_paper,reig2018spin}.
Indeed, both MgEr$_2$Se$_4$ \cite{reig2018spin} and CdEr$_2$Se$_4$ \cite{10_Lago_CdEr2Se4_heat_cap,18_Gao_CdEr2Se4_fennel_paper} have independently been identified as strong candidates for a classical spin ice state.
Ordered states have been observed in both MgYb$_2$X$_4$\cite{17_Higo_AYb2S4} and CdYb$_2$X$_4$\cite{17_Reotier_order_and_tunneling_in_CdYb2Se4} X $\in$ \{S, Se\}, but are notable for their highly renormalized moments and the existence of persistent spin dynamics at low temperatures\cite{17_Higo_AYb2S4,17_Reotier_order_and_tunneling_in_CdYb2Se4}.
Similar anomalous fluctuations are reported in CdHo$_2$S$_4$ \cite{15_Yaouanc_CdHo2S4} below a reported ordering transition, along with several features which draw parallels to the ``partially ordered'' pyrochlore system Tb$_2$Sn$_2$O$_7$\cite{07_rule_partial_order_Tb2Sn2O7,09_rule_Tb2Sn2O7}.
The presence of an ordered state and of a local moment size of $8~\mu_B$ in CdHo$_2$Se$_4$\cite{15_Yaouanc_CdHo2S4} are in direct contradiction to the singlet ground state predicted from magnetization measurements \cite{80_Ben-Dor_CdRE2S_series}.

In each of the above cases, the exact nature of the magnetic ground state is closely entwined with the local CEF environment of the constituent RE moments, as has been acknowledged on several occasions\cite{10_Lago_CdEr2Se4_heat_cap,reig2018spin,18_Gao_CdEr2Se4_fennel_paper,17_Higo_AYb2S4,15_Yaouanc_CdHo2S4,18_rau_Ba3Yb2Zn5O11}.
A necessary condition for realizing spin ice physics in MgEr$_2$Se$_4$\cite{reig2018spin} and CdEr$_2$Se$_4$\cite{10_Lago_CdEr2Se4_heat_cap,18_Gao_CdEr2Se4_fennel_paper} is the presence of a ground state Kramers doublet with Ising symmetry. The ordered states in MgYb$_2$X$_4$ and CdYb$_2$X$_4$ have been discussed in the context of frustrated anisotropic exchange models, in which the particular choice of CEF parameters can select from a variety of distinct ordered or spin liquid phases\cite{18_rau_Ba3Yb2Zn5O11}.
Material specific calculations predict the existence of several low energy CEF states in CdHo$_2$S$_4$\cite{18_Gao_CdEr2Se4_fennel_paper}, which draw even stronger parallels between this material and Tb$_2$Sn$_2$O$_7$ and lend special significance to the low-temperature fluctuations\cite{07_Molavian_virtual_CEF_excitations,05_mirebeau_LT_fluctuations_TbSnO,07_mirebeau_CEF_TbTiO_TbSnO}.

There thus exists a strong motivation for systematic and high precision measurements of the CEF Hamiltonian across this family of compounds.
Some early studies of cadmium spinels acquired this information through careful fits of magnetization data, however the stated results are broadly inconsistent with conclusions from modern studies\cite{74_Pokrzywnicki_CEF_CdYb2Se4,75_Pokrzywnicki_CEF_CdYb2Se4,88_pawlak_Mag_CEF_Yb,80_Ben-Dor_CdRE2S_series,77_pokrzywnicki_mag_tm2se3_CdTm2Se4}.
In the current paper, we instead measure crystal field excitations directly using inelastic neutron scattering (INS), and use fits of both the energy and intensity of observed transitions to determine the most likely CEF Hamiltonian consistent with the symmetry of the $Fd\bar{3}m$ space group.
This spectroscopic analysis is the de-facto choice for CEF measurements in rare earth systems\cite{12_Bertin_CEF_across_R2Ti2O7,99_siddharthan,champion2003er,00_rosenkranz} due to the method's precision and symmetry-driven modeling, which is largely insensitive to the presence of impurity phases, defects and other mechanisms which adversely affect bulk thermodynamic data.
In a recent publication, we showed how a similar analysis of INS data can be used to confirm the Ising-like effective spins in the material MgEr$_2$Se$_4$, and additionally identified several low-lying CEF excitations capable of inducing quantum fluctuations\cite{reig2018spin}. Below, we extend this analysis to three other closely related spinel selenide systems. In MgTm$_2$Se$_4$, we confirm the ground state is well characterized as a spin singlet, with the first excited state at $E = 0.876(16)$~meV -- low enough to thermally excite non-zero local moments at temperatures of only a few Kelvin.
In MgHo$_2$Se$_4$, we identify 10 separate CEF excitations, and determine an Ising-like ground state Kramers doublet with multiple low-lying excited states, drawing intriguing parallels to the 227 pyrochlores Tb$_2$Ti$_2$O$_7$ and Tb$_2$Sn$_2$O$_7$. Fits of MgYb$_2$Se$_4$ were underconstrained, but we determine a best fit Hamiltonian which is capable of reproducing both INS and magnetization data at a variety of fields. Compared to previous estimates\cite{17_Higo_AYb2S4,18_rau_Ba3Yb2Zn5O11}, our analysis is notable for the much stronger inferred easy-axis anisotropy. Results for each material are compared to measured magnetization data, and the implications for spin-spin interactions and magnetic ground states are discussed.

\begin{figure}[tbh]
\centering
\includegraphics[width=\columnwidth]{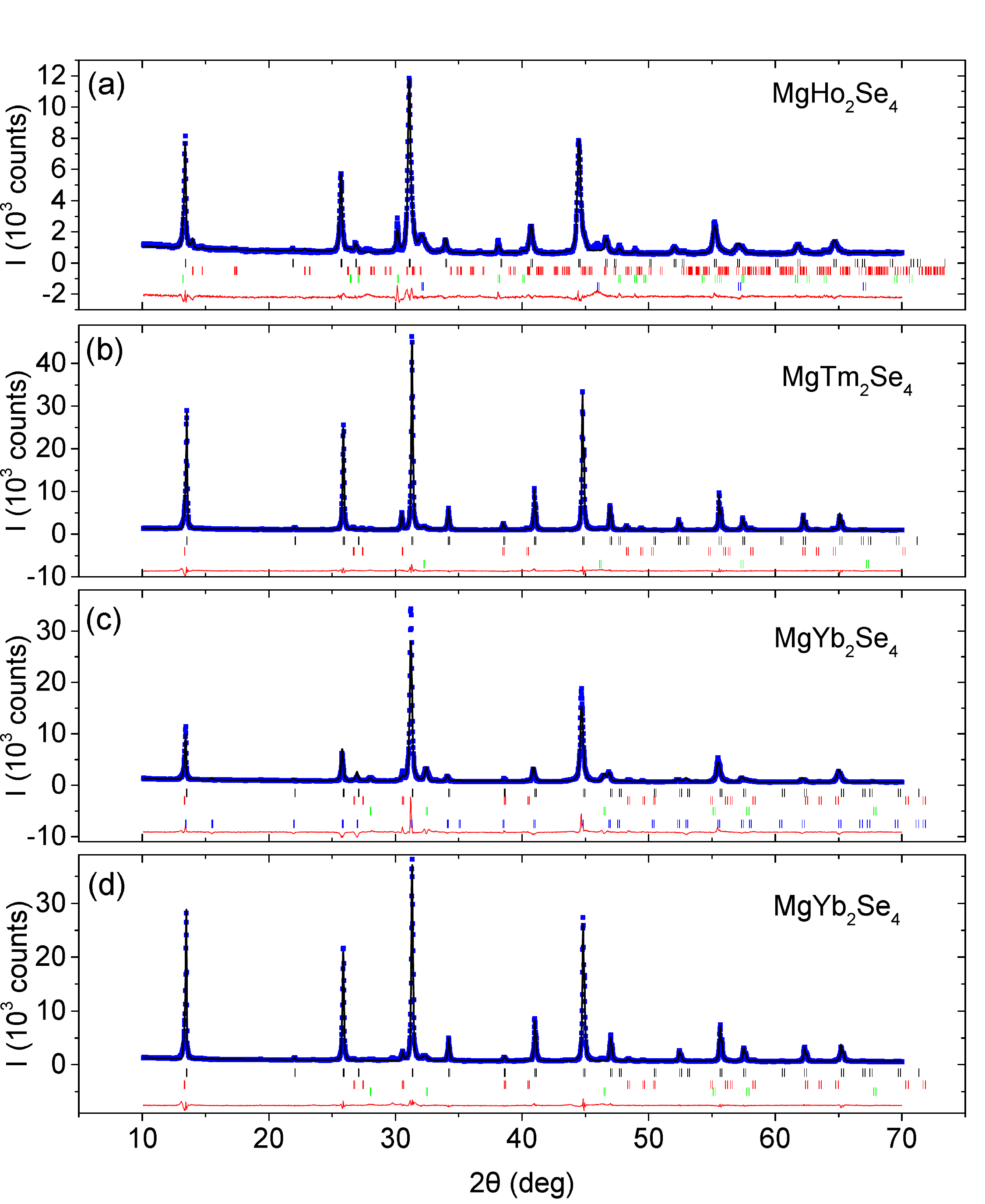}
\caption{Powder XRD plots of the measured MgRE$_2$Se$_4$ (RE = Ho, Tm and Yb from top to bottom, respectively) where data points are shown as blue dots, best fit refinement curves as black lines, and the difference curves are presented in red below the data. The two different RE = Yb plots are for the sample used for INS measurements (c) and magnetization (d). Tick marks below the data denote peaks of the majority phase and impurity phases, presented top to bottom in the same order as in Table~\ref{tab:impurity}.}
\label{fig:xrd}
\end{figure}

\section{Experimental methods}
Polycrystalline samples of RE spinel chalcogenides were prepared via a two-step solid state reaction at Illinois using the same method described in detail in Ref.~\onlinecite{reig2018spin}, and sample quality was confirmed via powder x-ray diffraction (XRD) using a PANalytical X'Pert$^3$ powder diffractometer at the Center of Nanophase Materials Science at Oak Ridge National Laboratory (ORNL).
INS measurements were performed using the SEQUOIA Fine-Resolution Fermi Chopper Spectrometer at the ORNL's Spallation Neutron Source.
Spectra were measured with a variety of initial neutron energies, $E_i$ and temperatures, $T$, as dictated by the relevant energy scales of the CEF transitions predicted by point charge calculations.
Specifically,  measurements were taken with $E_i =  6$, 11, 30, and 50~meV and $T =  5$ and 100~K for MgHo$_2$Se$_4$,  with $E_i = 30$, 50, and 100~meV and $T =  5$ and 100~K for MgTm$_2$Se$_4$, and with $E_i = 30$, 50, and 100~meV and $T = 5$ and 250~K for MgYb$_2$Se$_4$.
Magnetization measurements were taken on a Quantum Design MPMS3 instrument in the Illinois Materials Research Laboratory, utilizing the DC measurement mode.
Measurements were performed at temperatures of $T$ = 2, 5, 10, 20, and 40 K for all samples, with additional measurements at 80 and 120 K for MgHo$_2$Se$_4$.
Supplementary measurements were performed on MgYb$_2$Se$_4$ as a function of temperature at a constant field of 100 Oe, as laid out below.

\section{X-ray diffraction}
Figure~\ref{fig:xrd} shows the results of powder XRD measurements, along with solid curves representing the best refinements using the FULLPROF software suite\cite{rodriguez1993recent}.
Refinements assumed a majority phase with the $Fd\bar{3}m$ symmetry expected for a normal spinel structure, while accounting for the possibility of several common impurity phases.
The positions of associated Bragg peaks are indicated by sets of lines in Figure~\ref{fig:xrd}, with the majority phase indicated at the top in black and the impurity phases below in different colors.
The brightest reflections were reliably fit to the spinel MgRE$_2$Se$_4$, with the resulting cubic lattice parameters and fractional coordinate of the Se anions listed in Table~\ref{tab:struct_param}.
The values for lattice parameters are $\sim10\%$ larger than those typically observed in 227 pyrochlore oxides\cite{10_gardner} and, combined with a larger RE-anion distance,  result in a significantly lower energy scale for CEF excitations in the spinel selenides.
The fractional coordinate of the Se anions in the $Fd\bar{3}m$ space group quantify the trigonal distortion of RESe$_6$ octahedra, with measured values showing minimal deviation from the undistorted case at $x = 0.25$.
Accordingly, subsequent point charge calculations demonstrate that the dominant non-cubic contribution to the CEFs at the RE site comes from neighboring cations, and not distortions of the local chalcogen environment.
This observation is in conflict with previously used models of CEFs for these compounds\cite{74_Pokrzywnicki_CEF_CdYb2Se4,75_Pokrzywnicki_CEF_CdYb2Se4,17_Higo_AYb2S4}, and provides a further point of contrast between spinels and 227 pyrochlores.

The composition of the prepared samples varied with each synthesis, but all contained the same limited number of impurity phases. The exact distribution of phases in the large volume samples studied with INS are listed in Table~\ref{tab:impurity}, along with one higher purity sample of MgYb$_2$Se$_4$ which was prepared for follow-up studies of magnetization. In addition to the majority spinel phase, all samples investigated in this study had sizable fractions of  binary rare earth selenide compounds. This is consistent with the high vapor pressure of Mg, resulting in loss during reaction steps. All samples were further seen to contain between 3-8\% of the rare earth oxiselenide, which is consistent with the strong tendency for precursor metals to oxidize before forming selenium binaries.
The tendency towards metallicity and, in the case of YbSe the lack of local moments, minimize the contribution of the RE monoselenide (RESe) impurities to the INS spectrum \cite{64_reid_RESe}. The RE-sesquiselenides (RE$_2$Se$_3$)\cite{95_prokofiev_rare_earth_selenide_growth} and oxiselenides (RE$_2$O$_2$Se)\cite{72_quezel_oxyselenide_mag} are insulators with known structures and were accounted for in subsequent analysis. It is worth noting at this point, however, that the RE-oxiselenides have diffraction patterns which overlap substantially with peaks of the predicted spinel patterns, and have further been reported to have antiferromagnetic ordering transitions at temperatures below 5~K\cite{72_quezel_oxyselenide_mag}. The existence of previously unappreciated volume fractions of oxiselenide impurities is thus a leading contender to explain reports of unindexable magnetic Bragg peaks in a number of published  neutron powder diffraction studies of RE spinel chalcogenides\cite{reig2018spin,18_Gao_CdEr2Se4_fennel_paper,17_Reotier_order_and_tunneling_in_CdYb2Se4}.

\begin{table}[bht]

\begin{ruledtabular}
\begin{tabular}{ l l l }
Material & $a$ (\AA) & $x_{\text{Se}}$ \tabularnewline
\hline
MgHo$_2$Se$_4$ & 11.5508(2) & 0.2466(1) \tabularnewline
MgEr$_2$Se$_4$\cite{reig2018spin} & 11.5207(14) & 0.2456(9) \tabularnewline
MgTm$_2$Se$_4$ & 11.48493(5) & 0.24614(7) \tabularnewline
MgYb$_2$Se$_4$ & 11.45591(3) & 0.24595(8)\tabularnewline
\end{tabular}
\end{ruledtabular}
 \caption{Table listing the fit cubic lattice parameter ($a$), and the partial coordinate position of Se ($x_{\text{Se}}$) extracted from Rietveld refinements of XRD data. }.
 \label{tab:struct_param}
\end{table}

\begin{table*}[bht]

\begin{ruledtabular}
\begin{tabular}{ l l l l l l l l l l l l}
MgHo$_2$Se$_4$&( \tikz{\draw[line width=1.1pt]  (0,0.2) -- (0,0) ; } ) & 71.15( 0.68) &
MgTm$_2$Se$_4$&( \tikz{\draw[line width=1.1pt]  (0,0.2) -- (0,0) ; } ) & 90.31( 0.40) &
MgYb$_2$Se$_4$&( \tikz{\draw[line width=1.1pt]  (0,0.2) -- (0,0) ; } )& 68.11(0.86) &
MgYb$_2$Se$_4$&( \tikz{\draw[line width=1.1pt]  (0,0.2) -- (0,0) ; } )& 92.01( 0.52)
\tabularnewline

Ho$_2$Se$_3$&( \tikz{\draw[line width=1.1pt, red]  (0,0.2) -- (0,0) ; } ) & 8.84( 0.22) &
Tm$_2$O$_2$Se&( \tikz{\draw[line width=1.1pt, red]  (0,0.2) -- (0,0) ; } ) & 5.57( 0.06)&
Yb$_2$O$_2$Se&( \tikz{\draw[line width=1.1pt, red]  (0,0.2) -- (0,0) ; } ) & 3.92(0.28)&
Yb$_2$O$_2$Se&( \tikz{\draw[line width=1.1pt, red]  (0,0.2) -- (0,0) ; } ) & 3.02( 0.06)
\tabularnewline

Ho$_2$O$_2$Se&( \tikz{\draw[line width=1.1pt, green]  (0,0.2) -- (0,0) ; } ) &7.62( 0.14) &
TmSe&( \tikz{\draw[line width=1.1pt, green]  (0,0.2) -- (0,0) ; } ) & 4.12( 0.08) &
YbSe&( \tikz{\draw[line width=1.1pt, green]  (0,0.2) -- (0,0) ; } ) & 14.88(0.33) &
YbSe&( \tikz{\draw[line width=1.1pt, green]  (0,0.2) -- (0,0) ; } ) & 4.97( 0.13)
\tabularnewline

HoSe&( \tikz{\draw[line width=1.1pt, blue]  (0,0.2) -- (0,0) ; } ) & 12.38( 0.24) &
     &              &&
Yb$_{7.24}$Se$_8$&( \tikz{\draw[line width=1.1pt, blue]  (0,0.2) -- (0,0) ; } ) & 13.09(0.21) &
     &&
\end{tabular}
\end{ruledtabular}
 \caption{A list of all observed phases and their percent masses in the samples used in this paper. Next to each compound name is an example of the mark showing their peak positions in Fig.~\ref{fig:xrd}.}
 \label{tab:impurity}
\end{table*}

\begin{figure}[tbh]
\centering
\includegraphics[width=\columnwidth]{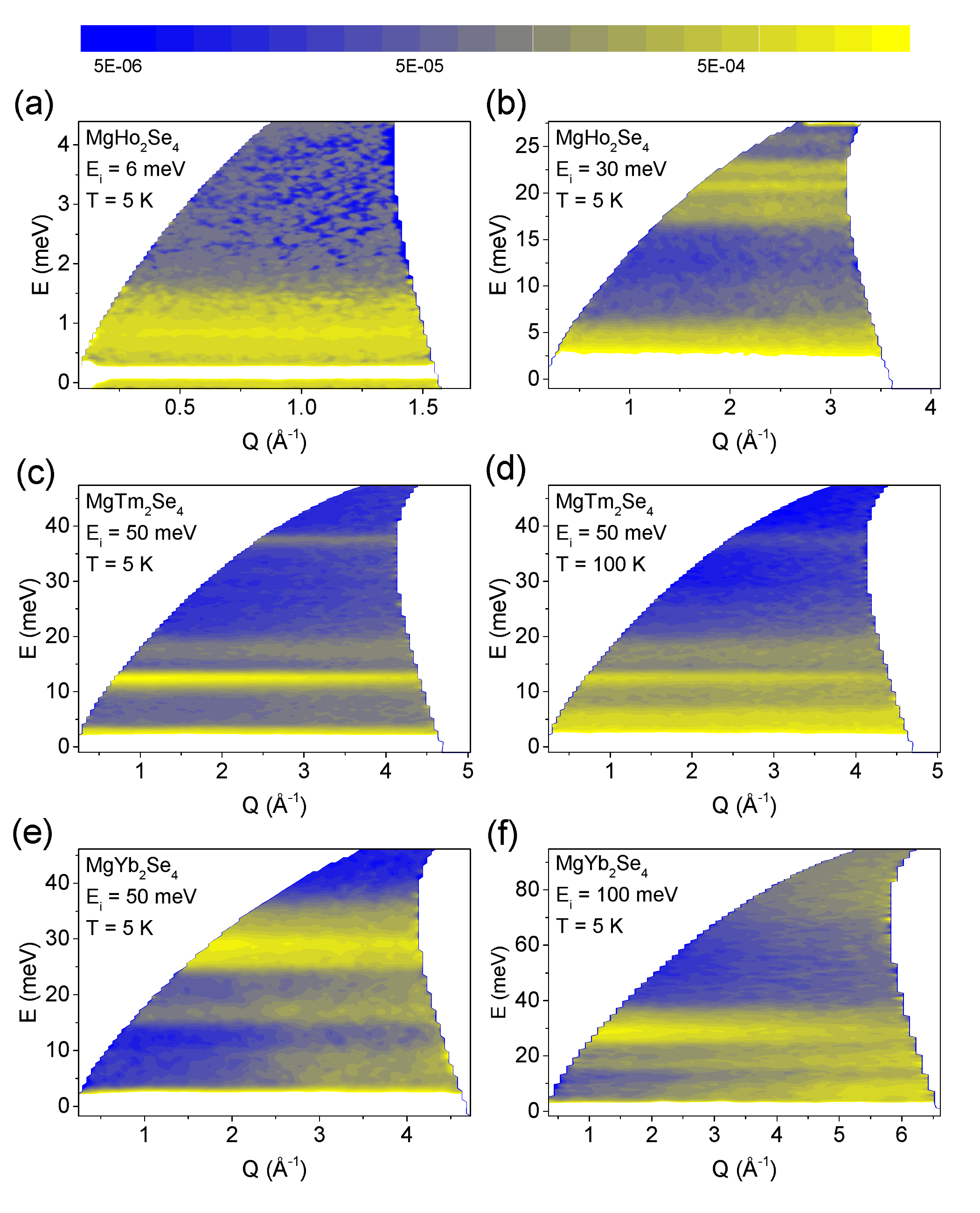}
\caption{Typical INS scattering spectra from powders of (a, b) MgHo$_2$Se$_4$  (c, d) MgTm$_2$Se$_4$, and (e, f) MgYb$_2$Se$_4$.}
\label{fig:SEQ_false_color}
\end{figure}

\begin{figure}[tbh]
\centering
\includegraphics[width=\columnwidth]{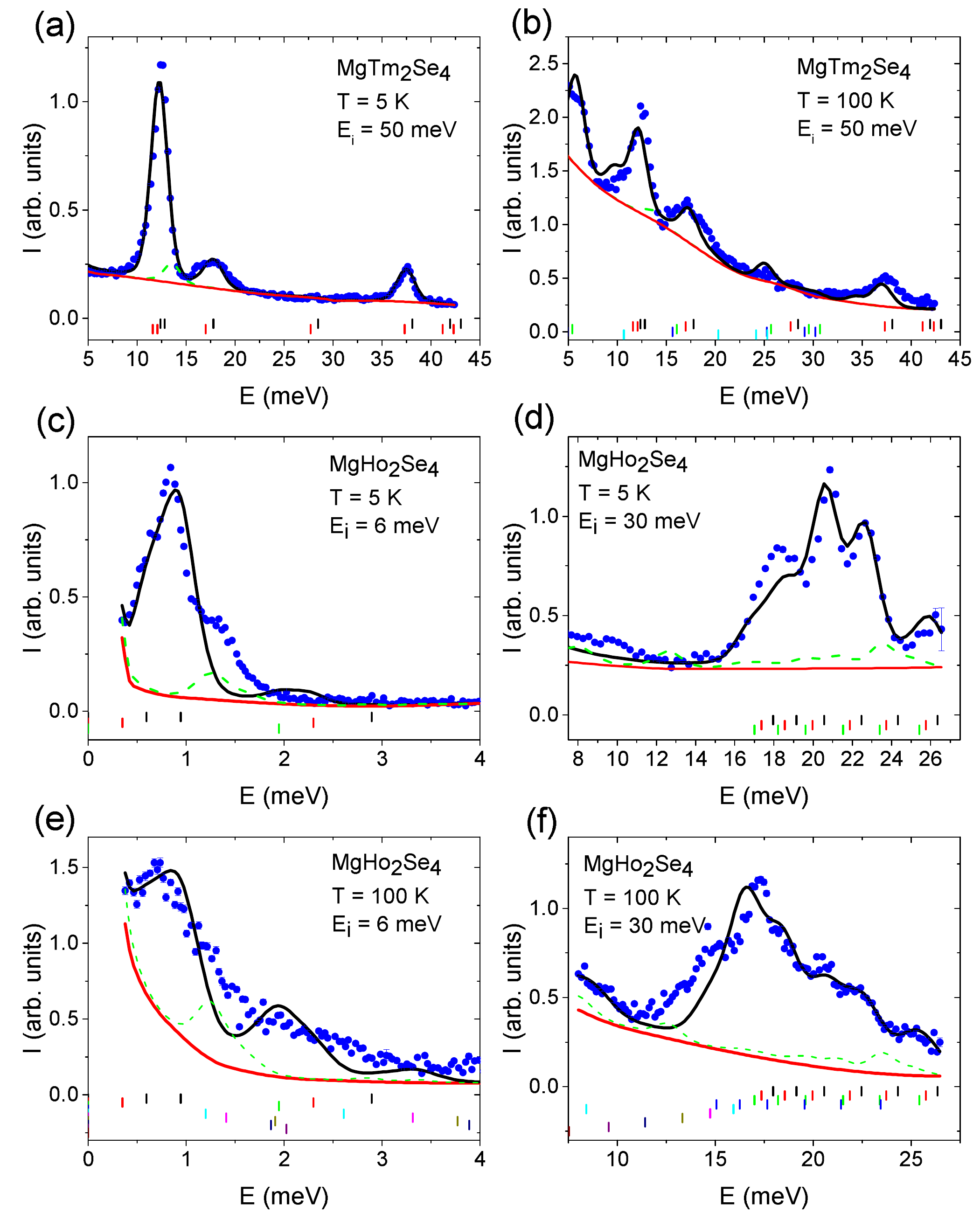}
\caption{ Constant-$Q$ cuts of MgTm$_2$Se$_4$ (a, b) and  MgHo$_2$Se$_4$ (c - f) data, with integration ranges and experimental conditions noted. The measured intensity is indicated by blue circles, the best fits are shown as solid black curves, the slowly-varying background contributions are indicated by solid red curves, and the dashed green curves depict possible contributions from impurity phases. Colored tick marks show the positions where CEF transitions lead to a peak with the black topmost marks representing excitations from the ground state, and each set below representing excitations from each following excited level.}
\label{fig:SEQ_ho}
\label{fig:SEQ_tm}
\end{figure}

\begin{figure}[tbh]
\centering
\includegraphics[width=\columnwidth]{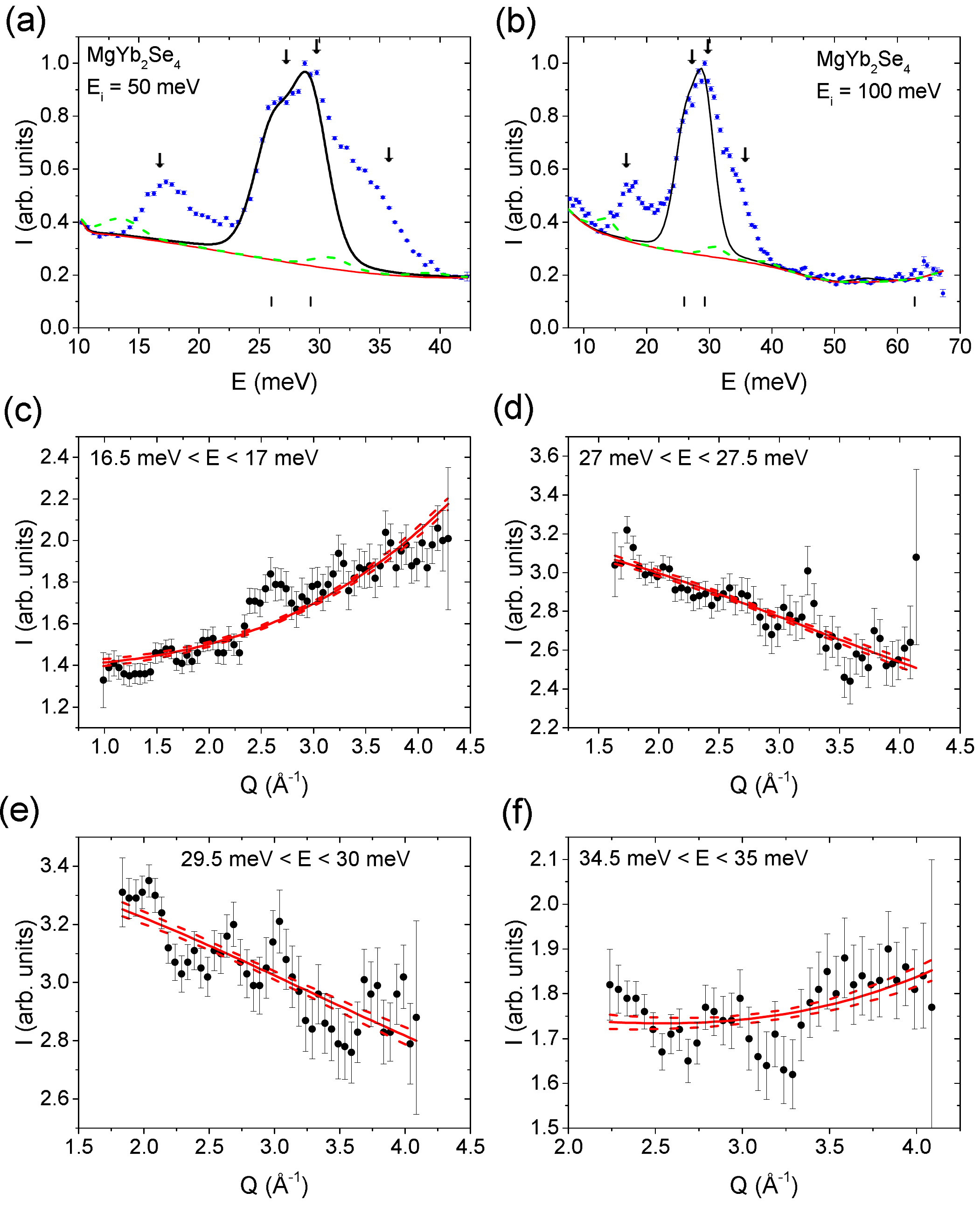}
\caption{(a,b) Constant-$Q$ cuts of MgYb$_2$Se$_4$ data taken with $T$ = 5$K$. The measured intensity is indicated by blue circles, the best fits are shown as solid black curves, the slowly-varying background contributions are indicated by solid red curves, and the dashed green curves depict possible contributions from impurity phases. Colored tick marks show the positions where CEF transitions lead to a peak. (c - f) Constant-E cuts of the MgYb$_2$Se$_4$ data with $E_i = 50$~meV and $T = 5$~K, integrated over listed energy ranges. The center of the integration range for each of these cuts is indicated by the arrows in (a) and (b).}
\label{fig:yb_cuts}
\end{figure}

\section{Inelastic neutron scattering}

Typical INS spectra for MgHo$_2$Se$_4$, MgTm$_2$Se$_4$ and MgYb$_2$Se$_4$ are shown in Fig.~\ref{fig:SEQ_false_color}, which for each material reveal the existence of multiple dispersionless modes at finite energy transfer.
The scattering intensity has contributions from both CEF transitions and phonons, in addition to various sources of background. The variation of the scattering intensity as a function of momentum transfer $(Q)$ was used to determine whether observed scattering modes originate from phonons or are magnetic. Each spectrum was measured with multiple incident neutron energies, as a means of separating intrinsic and spurious sources of scattering and to balance energy range and resolution.

Fig.~\ref{fig:SEQ_tm} and Fig.~\ref{fig:yb_cuts} show the variation of neutron scattering intensity versus energy, extracted from data in Fig.~\ref{fig:SEQ_false_color} by integrating over finite regions in $Q$ at positions chosen to maximize the available fit range at each incident neutron energy, $E_i$. Data is represented by blue dots, whereas solid curves represent the results of fits described in following sections. The solid red curves are estimates of the slowly varying contributions to background scattering, obtained by performing a cubic spline interpolation between points chosen away from obvious peaks. For all materials, the scattering above the slowly varying background takes the form of well defined peaks which are largely described by the CEF transitions laid out below.  Tick marks below the data show the energy positions of thermally accessible transitions between CEF states, described in more detail in the figure caption.

As shown in Fig.~\ref{fig:SEQ_tm}(a) and (b), the magnetic and phonon excitations are well-separated in the material MgTm$_2$Se$_4$ and therefore the identification of the CEF transitions is most straightforward.
The best fit curve shows excellent agreement with the data at both base temperature (5K) and at 100~K, where transitions from thermally populated levels contribute significantly to the scattering pattern. The INS data for MgHo$_2$Se$_4$ presented in Fig.~\ref{fig:SEQ_ho}(c)-(f) shows multiple overlapping peaks below 30 meV, but they are still clearly above background and mostly captured by the CEF fits.
The only exceptions are observed excesses of scattering at energies $E \approx 1.3$~meV and 16~meV, which we respectively associate with an impurity phase discussed below and with a phonon mode also seen in MgYb$_2$Se$_4$.

In MgYb$_2$Se$_4$, the CEF excitations overlap appreciably with dispersionless optical phonon modes, which mildly complicates analysis. The constant-E cuts for MgYb$_2$Se$_4$ shown in Fig.~\ref{fig:yb_cuts}(a) and (b) reveal four different modes -- a distinct peak near 17~meV, and three closely grouped peaks between 22 and 38~meV. Integrating over a finite energy range in the relevant spectra allows us to compare the $Q$-dependence of these excitations with the expectations for magnetic and phonon modes.
The cuts presented in Fig.~\ref{fig:yb_cuts}(c) and (f) clearly reveal the modes at $E = 16.75$~meV and 43.75~meV to be phonons, as the $Q$-dependence of the intensity varies as $I \propto Q^2$. On the other hand, the cuts shown in Fig.~\ref{fig:yb_cuts}(d) and (e) have $I \propto f(Q)^2$, where $f(Q)$ is the magnetic form factor for Yb$^{3+}$, and therefore these excitations are identified as CEF levels.
The phonon mode near 17~meV is consistent with the excess scattering in both MgTm$_2$Se$_4$ and MgHo$_2$Se$_4$ spectra at the same energy.  The two identified CEF excitations constitute two of three predicted transitions for MgYb$_2$Se$_4$ at $T = 5$~K, which is a $J = 7/2$ system\cite{kramers1930theorie}, though the absence of the third transition in the measured spectra is significant in that it places a upper bound on its scattering intensity.

\subsection{CEF model fitting}

The INS constant-Q cut data shown in Fig.~\ref{fig:SEQ_tm}(a)--(f) and Fig.~\ref{fig:yb_cuts}(a) and (b) was fit was fit using the Stevens operator approach, which considers only the ground state J-multiplet determined by Hund's rules.
This assumption is what is known as the LS coupling model \cite{16_ruminy_CEF227}, which has been demonstrated to be valid for rare earth atoms heavier than Dy\cite{18_gaudet,13_princep,00_rosenkranz} due to the large energy of the next J-multiplet \cite{53_elliott,64_freeman_SOC}.
The associated CEF Hamiltonian is given by
\begin{equation}
 H_{CEF} = \sum_{nm} B_{nm} \hat{O}_{nm},
\label{eq:CEF}
\end{equation}
where $B_{nm}$ are the CEF parameters and $\hat{O}_{nm}$ are Stevens operator equivalents\cite{stevens1952}, with the appropriate matrix elements for $\hat{O}_{nm}$ given by the software EasySpin \cite{06_stoll_easyspin}.

For the D$_{3d}$ point group symmetry of the RE-site, this reduces to:
\begin{multline}
 H_{CEF} = B_{20}\hat{O}_{20} + B_{40}\hat{O}_{40} + B_{43}\hat{O}_{43} + B_{60}\hat{O}_{60} \\
 + B_{63}\hat{O}_{63} + B_{66}\hat{O}_{66},
\label{eq:CEF_D3d}
\end{multline}
where we have chosen a quantization axis along the local $\langle111\rangle$ directions.

\begin{figure}[tbh]
\centering
\includegraphics[width=\columnwidth]{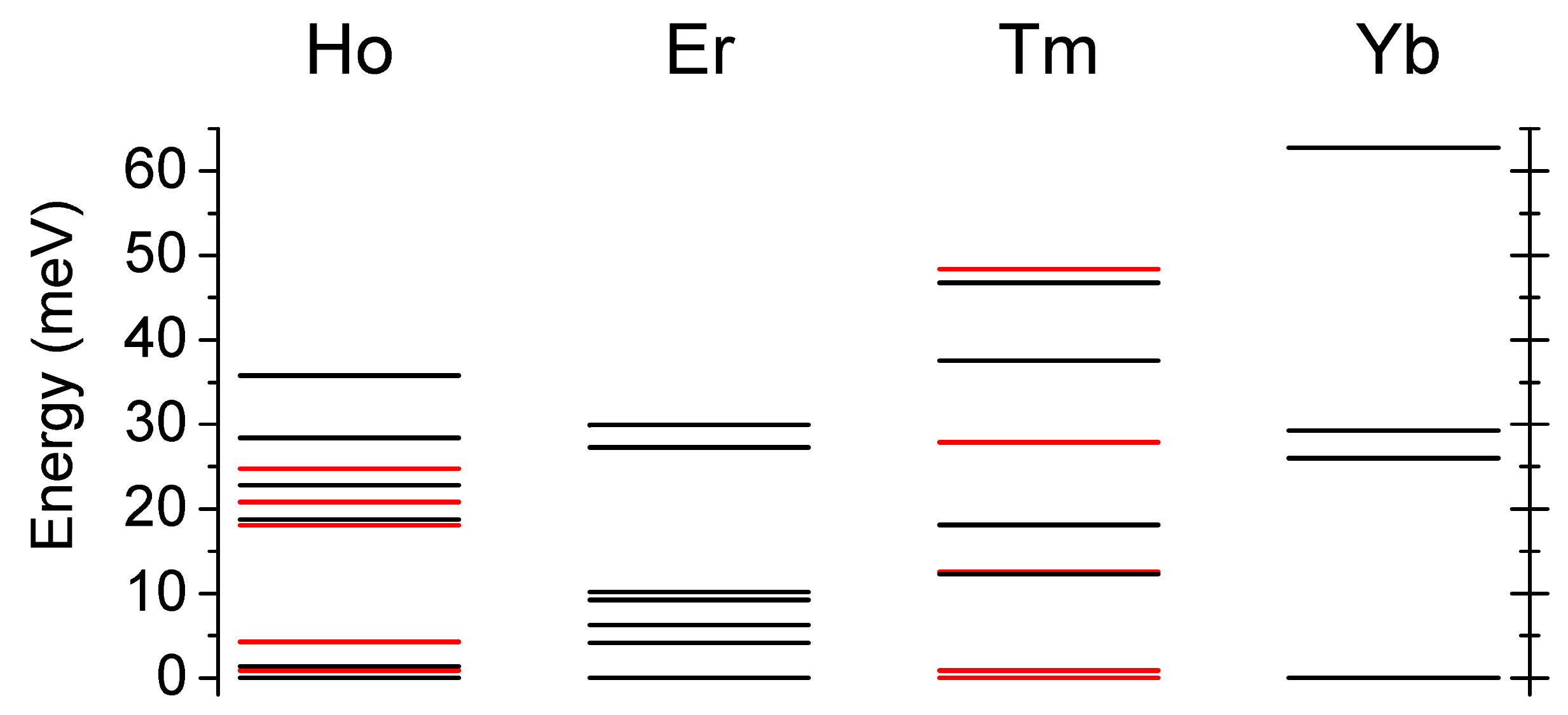}
\caption{A visualization of the CEF energy levels for MgRE$_2$Se$_4$ with RE = Ho, Er. Tm, and Yb. Doublets are shown in black and singlets are shown in red.}
\label{fig:CEF_level_visualization}
\end{figure}

For a given set of B$_{mn}$, CEF levels are found by direct diagonalization of Eq.~\ref{eq:CEF_D3d}, resulting in level schemes visualized in Fig.~\ref{fig:CEF_level_visualization}. The neutron scattering cross section of an excitation from the $i^{\text{th}}$ to the $j^{\text{th}}$ level is proportional to the matrix element given by
\begin{equation}
I_{ij} = \Sigma_\alpha \left|\bra{\psi_j} \hat{J}_\alpha \ket{\psi_i} \right|^2
\label{eq:CEF_intensity}
\end{equation}
where $\hat{J}_\alpha$ is the angular momentum operator in the $\alpha$ direction and $\ket{\psi_i}$ is the eigenket of the $i^{\text{th}}$ level. Total scattering intensity is modeled as the convolution of these matrix elements with a pseudo-Voigt instrument resolution function with a fitted width that was uniquely determined for each incident energy data set. Both excitations from the ground state and between excited states were considered, and each transition is  weighted with the appropriate Boltzmann factors at a given temperature.

The initial fit parameters were found by rescaling the published CEF parameters from our previous work on the material or MgEr$_2$Se$_4$\cite{reig2018spin} using
\begin{equation}
B_{nm} = \frac{\theta^{(n)} \langle r^n \rangle}{\theta^{(n)}_0 \langle r^n \rangle_0}
\left(\frac{a}{a_0} \right)^{-n+1} B^0_{nm},
\label{eq:CEF_scaling}
\end{equation}
as demonstrated in Refs.~\onlinecite{12_Bertin_CEF_across_R2Ti2O7} and \onlinecite{18_rau_Ba3Yb2Zn5O11}.
For these rescalings, the lattice parameters $a$ and $a_0$ are taken from our XRD fits,
we used $\left<r^n\right>$ found in  Ref.~\onlinecite{Forstreuter1997DFT}, and we used the Stevens parameters, $\theta^{(n)}$ defined in Ref.~\onlinecite{Hutchings1964227}. Subsequent fits then represent improvements over the rescaling predictions for CEF parameters, as they properly account for differences in local structure and covalency between individual materials.

\begin{table}
\begin{ruledtabular}
 \begin{tabular}{llllllll}
\multicolumn{7}{c}{MgHo$_2$Se$_4$}\tabularnewline
\multicolumn{7}{l}{$i = 0$~~( \tikz{\draw[line width=1.1pt]  (0,0.2) -- (0,0) ;} )
}\tabularnewline
\multicolumn{4}{c}{best fit} & \multicolumn{3}{c}{rescaled}\tabularnewline
$j$ & $E (meV)$ &$I_{5~K}$ &$ I_{100~K}$ & $E (meV)$ & $I_{5~K}$ & $ I_{100~K}$  \tabularnewline
\hline
 1  & 0.59 & 36.29 & 14.47 & 0.76 & 51.61& 20.78 \tabularnewline
 2  & 0.95 & 52.00 & 20.74 & 0.87 & 35.71& 14.38 \tabularnewline
 3  & 2.88 & 0.30 & 0.12 & 2.69 & 0.81& 0.32 \tabularnewline
 4  & 17.74 & 0.64 & 0.26 & 20.02 & 0.61& 0.25 \tabularnewline
 5  & 19.20 & 0.69 & 0.28 & 20.29 & 0.54& 0.22 \tabularnewline
 6  & 20.71 & 4.48 & 1.79 & 21.92 & 4.59& 1.85 \tabularnewline
 7  & 22.56 & 2.76 & 1.10 & 23.54 & 3.15& 1.27 \tabularnewline
 8  & 24.27 & 0.05 & 0.02 & 26.07 & 0.00& 0.00 \tabularnewline
 9  & 26.08 & 0.74 & 0.29 & 27.63 & 0.34& 0.14 \tabularnewline
 10 & 34.87 & 0.09 & 0.03 & 31.54 & 0.13& 0.05 \tabularnewline
\hline
\tabularnewline
\multicolumn{7}{l}{ $i = 1$ ~~( \tikz{\draw[line width=1.1pt, red]  (0,0.2) -- (0,0) ;} )
}\tabularnewline
\hline
2   & 0.35 & 0.01 & 0.02 & 0.11 & 0.04& 0.09 \tabularnewline
3   & 2.29 & 1.93 & 2.84 & 1.93 & 5.40& 11.70 \tabularnewline
4   & 17.15 & 0.00 & 0.00 & 19.26 & 0.93& 2.02 \tabularnewline
5   & 18.61 & 1.14 & 1.67 & 19.53 & 0.32& 0.70 \tabularnewline
6   & 20.12 & 0.07 & 0.10 & 21.16 & 0.01& 0.03 \tabularnewline
7   & 21.97 & 0.07 & 0.10 & 22.78 & 0.48& 1.04 \tabularnewline
8   & 23.68 & 0.05 & 0.08 & 25.31 & 0.03& 0.06 \tabularnewline
9   & 25.49 & 0.15 & 0.22 & 26.87 & 0.13& 0.29 \tabularnewline
10  & 34.28 & 0.06 & 0.08 & 30.78 & 0.03& 0.07 \tabularnewline
\hline
\tabularnewline
\multicolumn{7}{l}{ $i = 2$ ~~ ( \tikz{\draw[line width=1.1pt, green]  (0,0.2) -- (0,0) ;} )
}\tabularnewline
\hline
3   & 1.94 & 3.44 & 11.02 & 1.82 & 0.98& 2.70 \tabularnewline
4   & 16.80 & 0.74 & 2.38 & 19.15 & 0.00& 0.00 \tabularnewline
5   & 18.26 & 0.22 & 0.69 & 19.41 & 0.71& 1.97 \tabularnewline
6   & 19.77 & 0.01 & 0.04 & 21.05 & 0.07& 0.21 \tabularnewline
7   & 21.61 & 0.21 & 0.66 & 22.66 & 0.04& 0.12 \tabularnewline
8   & 23.32 & 0.03 & 0.10 & 25.20 & 0.01& 0.03 \tabularnewline
9   & 25.13 & 0.18 & 0.57 & 26.75 & 0.03& 0.10 \tabularnewline
10  & 33.93 & 0.02 & 0.06 & 30.67 & 0.01& 0.04 \tabularnewline
 \end{tabular}
\end{ruledtabular}
\caption{A list of the expected energies and neutron scattering cross sections of transitions from the the $i^{th}$ to $j^{th}$ CEF levels in MgHo$_2$Se$_4$. Listed are both expectations from fitted $B_{mn}$ presented in Table~\ref{tab:BNM_energies}, and from rescaling the CEF potential of MgEr$_2$Se$_4$.}
\label{tab:intensities_ho}
\end{table}

\begin{table}
\begin{ruledtabular}
 \begin{tabular}{llllllll}
\multicolumn{7}{c}{MgTm$_2$Se$_4$}\tabularnewline
\multicolumn{7}{l}{ $i = 0$~~( \tikz{\draw[line width=1.1pt]  (0,0.2) -- (0,0) ;} )
}\tabularnewline
\multicolumn{4}{c}{best fit} & \multicolumn{3}{c}{rescaled}\tabularnewline
$j$ & $E (meV)$ &$I_{5~K}$ &$ I_{100~K}$ & $E (meV)$ & $I_{5~K}$ & $ I_{100~K}$  \tabularnewline
\hline
1   & 0.88 & 28.92 & 11.12 & 0.57 & 25.62& 12.27 \tabularnewline
2   & 12.26 & 6.56 & 2.52 & 15.49 & 5.67& 2.71 \tabularnewline
3   & 12.55 & 0.00 & 0.00 & 16.66 & 0.00& 0.00 \tabularnewline
4   & 18.12 & 0.64 & 0.25 & 20.34 & 0.54& 0.26 \tabularnewline
5   & 27.92 & 0.01 & 0.00 & 33.31 & 0.19& 0.09 \tabularnewline
6   & 37.57 & 0.97 & 0.37 & 42.51 & 1.09& 0.52 \tabularnewline
7   & 46.76 & 0.04 & 0.02 & 46.04 & 0.00& 0.00 \tabularnewline
8   & 48.41 & 0.00 & 0.00 & 47.18 & 0.00& 0.00 \tabularnewline
\hline
\tabularnewline
\multicolumn{7}{l}{ $i = 1$ ~~ ( \tikz{\draw[line width=1.1pt, red]  (0,0.2) -- (0,0) ;)} )
   }\tabularnewline
\hline
 2  & 11.38 & 0.11 & 0.28 & 14.92 & 0.24& 0.40 \tabularnewline
 3  & 11.68 & 0.33 & 0.87 & 16.09 & 0.44& 0.73 \tabularnewline
 4  & 17.25 & 0.40 & 1.07 & 19.77 & 0.97& 1.63 \tabularnewline
 5  & 27.04 & 0.00 & 0.00 & 32.74 & 0.00& 0.00 \tabularnewline
 6  & 36.70 & 0.21 & 0.56 & 41.94 & 0.18& 0.30 \tabularnewline
 7  & 45.88 & 0.02 & 0.07 & 45.47 & 0.01& 0.01 \tabularnewline
 8  & 47.54 & 0.00 & 0.00 & 46.62 & 0.18& 0.30 \tabularnewline
\hline
\tabularnewline
\multicolumn{7}{l}{ $i = 2$~~( \tikz{\draw[line width=1.1pt, green]  (0,0.2) -- (0,0) ;)} )
}\tabularnewline
\hline
3   & 0.29 & 0.00 & 2.14 & 1.17 & 0.00& 1.61 \tabularnewline
4   & 5.86 & 0.00 & 3.59 & 4.85 & 0.00& 2.79 \tabularnewline
5   & 15.66 & 0.00 & 0.00 & 17.82 & 0.00& 0.01 \tabularnewline
6   & 25.31 & 0.00 & 0.21 & 27.02 & 0.00& 0.10 \tabularnewline
7   & 34.50 & 0.00 & 0.26 & 30.55 & 0.00& 0.01 \tabularnewline
8   & 36.16 & 0.00 & 0.00 & 31.69 & 0.00& 0.21 \tabularnewline

 \end{tabular}
\end{ruledtabular}
\caption{A list of the expected energies and neutron scattering cross sections of transitions from the the $i^{th}$ to $j^{th}$ CEF levels in MgTm$_2$Se$_4$. Listed are both expectations from fitted $B_{mn}$ presented in Table~\ref{tab:BNM_energies}, and from rescaling the CEF potential of MgEr$_2$Se$_4$.}
\label{tab:intensities_tm}
\end{table}

\begin{table}
\begin{ruledtabular}
\begin{tabular}{llllllll}
\multicolumn{7}{c}{MgYb$_2$Se$_4$}\tabularnewline
\multicolumn{7}{l}{$i = 0$~~( \tikz{\draw[line width=1.1pt]  (0,0.2) -- (0,0) ; } )
}\tabularnewline
\multicolumn{4}{c}{best fit} & \multicolumn{3}{c}{rescaled}\tabularnewline
$j$ & $E (meV)$ &$I_{5~K}$ &$ I_{100~K}$ & $E (meV)$ & $I_{5~K}$ & $ I_{100~K}$  \tabularnewline
\hline
1   & 26.01 & 4.10 & 3.78 & 19.02 & 5.73& 4.98 \tabularnewline
2   & 29.13 & 5.99 & 5.52 & 28.28 & 5.70& 4.96 \tabularnewline
3   & 54.95 & 0.13 & 0.12 & 54.12 & 0.07& 0.06 \tabularnewline
\hline
\tabularnewline
\multicolumn{7}{l}{ $i = 1$}\tabularnewline
\hline
  2 & 3.12 & 0.00 & 0.04 & 9.26 & 0.00& 0.17 \tabularnewline
  3 & 28.93 & 0.00 & 0.07 & 35.10 & 0.00& 0.24 \tabularnewline
\hline
\tabularnewline
\multicolumn{7}{l}{ $i = 2$ }\tabularnewline
\hline
  3 & 25.81 & 0.00 & 0.20 & 25.84 & 0.00& 0.20 \tabularnewline

 \end{tabular}
\end{ruledtabular}
\caption{A list of the expected neutron scattering cross sections of transitions from the first 3 CEF levels in MgYb$_2$Se$_4$, given by the best fit of the data, as well as what is expected from rescaling the CEF potential of MgEr$_2$Se$_4$.}
\label{tab:intensities_yb}
\end{table}

\begin{table*}[bht]
\begin{ruledtabular}
   \begin{tabular}{ l l l l l }
       & MgHo$_2$Se$_4$ & MgEr$_2$Se$_4$\cite{reig2018spin} & MgTm$_2$Se$_4$ & MgYb$_2$Se$_4$
      \tabularnewline
    \hline
     \tabularnewline
     B$_{20}$
     & $ 7.10(78)\cdot10^{-2} $ 
     & $-4.214(63)\cdot10^{-2}$ 
     & $-0.1707(5)$             
     & $ -0.507(61) $           
 \tabularnewline
     B$_{40}$
     & $ 5.52(24)\cdot10^{-4} $  
     & $ -6.036(30)\cdot10^{-4}$ 
     & $-1.880(6)\cdot10^{-3}$   
     & $ 2.89(4)\cdot10^{-2} $   
 \tabularnewline
     B$_{43}$
     & $ 9.07(13)\cdot10^{-3} $   
     & $ -1.3565(67)\cdot10^{-2} $
     & $-4.533(18)\cdot10^{-2}$   
     & $ 0.310(50) $              
 \tabularnewline
     B$_{60}$
     & $ -2.44(18)\cdot10^{-6} $  
     & $ 3.264(16)\cdot10^{-6} $  
     & $-4.59(11)\cdot10^{-6} $   
     & $ 1.98\cdot10^{-4} $   
 \tabularnewline
     B$_{63}$
     & $ 2.94(83)\cdot10^{-5} $   
     & $ -3.791(75)\cdot10^{-5} $ 
     & $ 2.48(2)\cdot10^{-4} $    
     & $ -2.30\cdot10^{-3} $  
 \tabularnewline
     B$_{66}$
     & $ -2.09(17)\cdot10^{-5} $  
     & $ 2.194(65)\cdot10^{-5} $  
     & $-1.441(59)\cdot10^{-4}$   
     & $ 1.33\cdot10^{-3} $   
 \tabularnewline
\end{tabular}
\end{ruledtabular}
 \caption{The CEF parameters (in meV) of the compounds MgRE$_2$Se$_4$ for RE = Ho, Tm and Yb from the best fits of the INS data shown in this paper, and results for RE = Er taken from our previous paper\cite{reig2018spin}.
 }
 \label{tab:BNM_energies}
\end{table*}

In Tables~\ref{tab:intensities_ho} ,~\ref{tab:intensities_tm} and ~\ref{tab:intensities_yb}, we list the energies and predicted neutron scattering intensities of relevant excitations calculated using CEF parameters from both the scaling analysis and best fits of neutron data, discussed below. The positions of these transitions are indicated in Figs.~\ref{fig:SEQ_tm} and \ref{fig:yb_cuts} with vertical tick marks.

For both MgHo$_2$Se$_4$ and MgTm$_2$Se$_4$, several transitions contribute to each of the peaks observed in constant-Q cuts of scattering data, though the intensity was overwhelmingly dominated by excitations out of the ground state. In order to access more transitions by thermal population of excited levels, we also include measurements at $T = 100$~K.
Refinements of CEF parameters were performed via a global least squares minimization routine using the constant-Q cuts presented in Fig.~\ref{fig:SEQ_ho}(a)-(f) and the predicted scattering intensity from all CEF transitions expected in the measured temperature range. The best fits are shown as solid red curves in these figures, and with few exceptions reproduce both the magnitude and position of all major peaks while predicting no scattering intensity that was not observed by experiment.

For MgYb$_2$Se$_4$, only the ground state CEF level has appreciable occupation at temperatures below 200 K, simplifying the magnetic spectrum.
However, the strong overlap between CEF and phonon peaks makes the above procedure untenable, as it fits raw neutron intensity and associates all non-background scattering with the Hamiltonian in Eq.~\ref{eq:CEF_scaling}.
Instead, the constant-Q cut data in Fig.~\ref{fig:yb_cuts}(a) and (b) were fit to multiple pseudo-Voigt peaks shown as solid lines, with resulting peak intensities listed in Table~\ref{tab:intensities_yb}.
The refinement of CEF parameters was subsequently performed by consideration of these fitted energies and intensities. To deal with the underconstrained nature of fitting 6 CEF parameters to only 4 pieces of information, we fixed the $B_{nm}$ values for $n = 6$ to the initial rescaled values and only refined parameters $B_{20}$, $B_{40}$, and $B_{43}$.
We subsequently verified that varying the parameters  $B_{60}$, $ B_{63}$ and $ B_{66}$ had minimal impact on the predicted neutron peak intensity and associated analysis.

The $B_{nm}$ parameters resulting from fits are given in Table~\ref{tab:BNM_energies}, along with uncertainties. For MgHo$_2$Se$_4$ and MgTm$_2$Se$_4$, uncertainties are determined by stepping out in one direction in parameter space while continually optimizing other parameters, until the reduced $\chi^2$ is increased by one. For the MgYb$_2$Se$_4$ fit, we again kept the $B_{nm}| n = 6$ values fixed when finding uncertainties. The full implications of these fitted parameters for the CEF levels and low-temperature effective spin systems of the three materials are laid out more fully in the following sections.

\subsection{Potential effect of impurities}

To consider the potential contribution to the CEF signal from impurity phases, we modeled the expected CEF scheme and the associated inelastic neutron scattering of relevant sesquiselenide and oxiselenide phases using a simple point charge model. For these, we assumed perfect ionic bonding, included all ions out to 7.5~\AA, and used structures taken from the above XRD refinements. The potential was calculated in a tesseral harmonic expansion $\gamma_{nm}$. For the cosine ($m \geq 0$) and sine ($m < 0$) components of the tesseral harmonics, we got the coefficients of the tesseral harmonics in Cartesian coordinates from Ref.~\onlinecite{61_prather_tesseral_coeff}. The CEF parameters are calculated as $B_{nm} = \langle r^n \rangle \theta_n \left(1-\sigma_n \right) \gamma_{nm}$, where $\langle r^n \rangle$ is the radius term, and $\sigma_n$ is the shielding parameter; both values are taken from calculations in Ref.~\onlinecite{98_edvardsson_CEF_shield}. The CEF Hamiltonian is then constructed using Eq.~\ref{eq:CEF}, and predicted neutron intensities are calculated as laid out above. For the calculations, we used the software EasySpin \cite{06_stoll_easyspin} to generate the matrix elements of the Stevens operators.

The calculated scattering from the CEF levels is scaled according to molar fraction of the ion in the sample and plotted in all of the constant-$Q$ cuts shown in Fig.~\ref{fig:yb_cuts}(a)--(b) and Fig.~\ref{fig:SEQ_ho}(a)--(d) as solid green curves. Similarly calculated CEF parameters \cite{98_edvardsson_CEF_shield,72_Gupta_sheilding_RE_HF_calc} are known to reproduce experimental values within about 20\%\cite{66_Blok_CEF_shielding_RE}, and are sufficient to reproduce the general shape and integrated intensity of peaks in neutron scattering spectra. Within these bounds on uncertainty, inspection of the calculated spectra can potentially explain excess scattering in the MgHo$_2$Se$_4$ spectra at 1.3~meV and 18~meV, and may overlap with peaks in MgYb$_2$Se$_4$ at 17~meV and 35~meV. Overall however, the energies of CEF levels from the impurity phases seem to be well removed from levels associated with the majority phases, and are significantly less intense. We thus conclude that excitations from impurities have a minimal effect on the fits of CEF levels laid out above.

\subsection{Results and interpretation}

In addition to producing the energy level schemes presented in Fig.~\ref{fig:CEF_level_visualization}, the fitted CEF parameters in Table~\ref{tab:BNM_energies} were used to calculate the associated eigenkets and, in particular, the ground state wavefunctions, which determine the size and anisotropy of moments in the low-temperature effective spin states.
In Table~\ref{tab:wavefunctions}, we list the resulting ground state wavefunctions for the three magnesium spinel compounds investigated in this paper, along with our previously determined results on the material MgEr$_2$Se$_4$\cite{reig2018spin}, included for comparison. The wavefunctions of degenerate doublets were determined with a small guide field artificially applied along the $\langle 111 \rangle$ direction.

With no further analysis, one can immediately see that MgTm$_2$Se$_4$ has a ground state singlet with no net moment, in line with the previous conclusions of Ref.~\onlinecite{80_Ben-Dor_CdRE2S_series}. For the other systems, the ground state is a doublet, which we can use to define a pseudo-spin-$\frac{1}{2}$ with effective up and down states (denoted $\ket{+} \text{and} \ket{-}$). With these states, we use
\begin{equation}
\frac{1}{2} g_{ii} \sigma_i = \begin{bmatrix}
       \bra{+} j_i \ket{+} & \bra{+} j_i \ket{-} \\
       \bra{-} j_i \ket{+} & \bra{-} j_i \ket{-}
       \end{bmatrix}
\end{equation}
to find the component of the moment parallel and perpendicular to the local $\langle 111\rangle$ directions. The results of these calculations are displayed in Table \ref{tab:g-factors}, where $\sigma_i$ are the Pauli matrices and $g_{zz}$ and $g_{xx}$ define $g_\parallel$ and $g_\perp$, respectively.

These values can be used to comment on the anisotropy of the effective spins. As an example, our previously determined results for MgEr$_2$Se$_4$ show $g_{\perp}=0$, indicating that material has fully Ising moments\cite{reig2018spin}. The current results imply that the moments in MgHo$_2$Se$_4$ also have perfect Ising symmetry, though one might expect deviations from this conclusion should one include spin-spin interactions which have the capacity to mix the CEF transitions.

For MgYb$_2$Se$_4$, we find $g_\parallel = 5.188(79)$ and $g_\perp = 0.923(85)$, implying an effective spin with strong anisotropy along the $\langle 111 \rangle$ direction while falling far short of the Ising limit. These values imply significantly more anisotropy than the values of $g_\parallel = 3.564$ and $g_\perp = 2.204 $ obtained from rescaling the CEF parameters from our previous MgEr$_2$Se$_4$ results\cite{reig2018spin,18_rau_Ba3Yb2Zn5O11}, and are even farther removed from reports of nearly isotropic spins determined from fits of inverse magnetic susceptibility curves\cite{17_Higo_AYb2S4,88_pawlak_Mag_CEF_Yb}. Comparing the CEF parameters from these and the current study, the starkest contrast lay in the signs of $B_{43}$ and $B_{63}$ parameters and the magnitude of $B_{20}$.
For MgYb$_2$Se$_4$, the parameters $B_{6m} | m = 0,~3,~6$ have little consequence on the relative sizes of $g_\parallel$ and $g_\perp$, and on the goodness of the fit to our data, however both $B_{20}$ and the ratio between $B_{40}$ and $B_{43}$ are strongly associated with trigonal fields and thus the tendency of moments to point in the $\langle 111 \rangle$ directions.
The magnetic susceptibility studies\cite{88_pawlak_Mag_CEF_Yb,17_Higo_AYb2S4} underestimate the $B_{20}$ parameter and, more consequentially, fix the ratio between $B_{40}$ and $B_{43}$ to that expected in a perfect octahedral environment. This hugely underestimates the contribution to the potential from the next nearest neighbor atoms, and is likely responsible for the resultant underestimation of the anisotropy of the Yb$^{3+}$ ions.

Finally, it is worth noting that the presence of optical phonons in several locations at the same energies as CEF levels raises the possibility that vibronic bound states might exist in these compounds. Vibronic bound states form as the result of strong magnetoelastic coupling, which produces significant hybridization/entanglement of CEF excitations and phonons\cite{84_Thalmeier_CEF-phonon_theory}, and have recently been observed in the related 227 pyrochlore compounds Ho$_2$Ti$_2$O$_7$\cite{18_gaudet_vibronic_states} and Tb$_2$Ti$_2$O$_7$\cite{16_ruminy_CEF227}.
These bound states appear in INS spectra as a splitting of either CEF or phonon excitations, resulting in scattering from new modes with unusual momentum and temperature dependences\cite{03_Loewenhaupt_CEF-phonon}. Of note in the current materials is the relatively intense optical phonon at E = 17 meV, which is close to the 17.7 meV CEF mode in MgHo$_2$Se$_4$ and 18 meV CEF mode in MgTm$_2$Se$_4$, as well as a possible 34 meV phonon mode which is near a 29.1 meV CEF mode in MgYb$_2$Se$_4$. Though initial inspection of our data reveals no clear evidence for vibronic modes, we suggest that the above materials and energies may be promising areas to search for bound states with follow-up higher resolution or polarized neutron scattering measurements.

\begin{table*}
\begin{ruledtabular}
 \begin{tabular}{l l l l}
 \tabularnewline
 \hline
 Ho & $\ket{\psi_0^\pm}  $ &=& $ 0.323(66)\ket{\pm7} \mp 0.711(11)\ket{\pm4} + 0.0797(45)\ket{\pm1} \mp 0.317(51)\ket{\mp2} - 0.516(17)\ket{\mp5} \mp 0.131(42)\ket{\mp8}$
 \tabularnewline
 Er\cite{reig2018spin} & $\ket{\psi_0^\pm}  $ &=& $ \pm0.9165(7)\ket{\pm15/2} + 0.3600(11)\ket{\pm9/2} \pm 0.1581(16)\ket{\pm3/2} - 0.0731(15)\ket{\mp3/2} \pm 0.0036(7)\ket{\mp9/2} + $
 \tabularnewline &  & & $ 0.0034(14)\ket{\mp15/2} $
 \tabularnewline
 Tm & $\ket{\psi_0} $ &=& $ 0.6700(5)\ket{6} + 0.1468(7)\ket{3} + 0.2431(19)\ket{0} - 0.1468(7)\ket{-3} + 0.6700(5)\ket{-6}$
 \tabularnewline
 Yb & $\ket{\psi_0^\pm}  $ &=& $ - 0.9684(48) \ket{\pm5/2} \pm 0.218(16)\ket{\mp 1/2} + 0.1204(99)\ket{\mp7/2} $
 \end{tabular}
\end{ruledtabular}
\caption{Calculated ground state wavefunctions for MgRE$_2$Se$_4$ with RE = Ho, Er, Tm, and Yb.}
\label{tab:wavefunctions}
\end{table*}

\begin{table}
\begin{ruledtabular}
 \begin{tabular}{l c c}
  & $g_\parallel$ & $g_\perp$ \tabularnewline
  \hline
 Ho & $2.72(46)$ & $0.000(63)$
 \tabularnewline
 Er\cite{reig2018spin} & $16.591(12)$ & $0$
 \tabularnewline
 Tm & $0$ & $0$
 \tabularnewline
 Yb & $5.188(79)$ & $0.923(85)$
 \end{tabular}
\end{ruledtabular}
\caption{Calculated values for the g-factors of the ground state of the rare earth ions in MgRE$_2$Se$_4$, relative to local  $\langle 111 \rangle$ directions.}
\label{tab:g-factors}
\end{table}

\begin{figure*}[tbh]
\includegraphics[width=\linewidth]{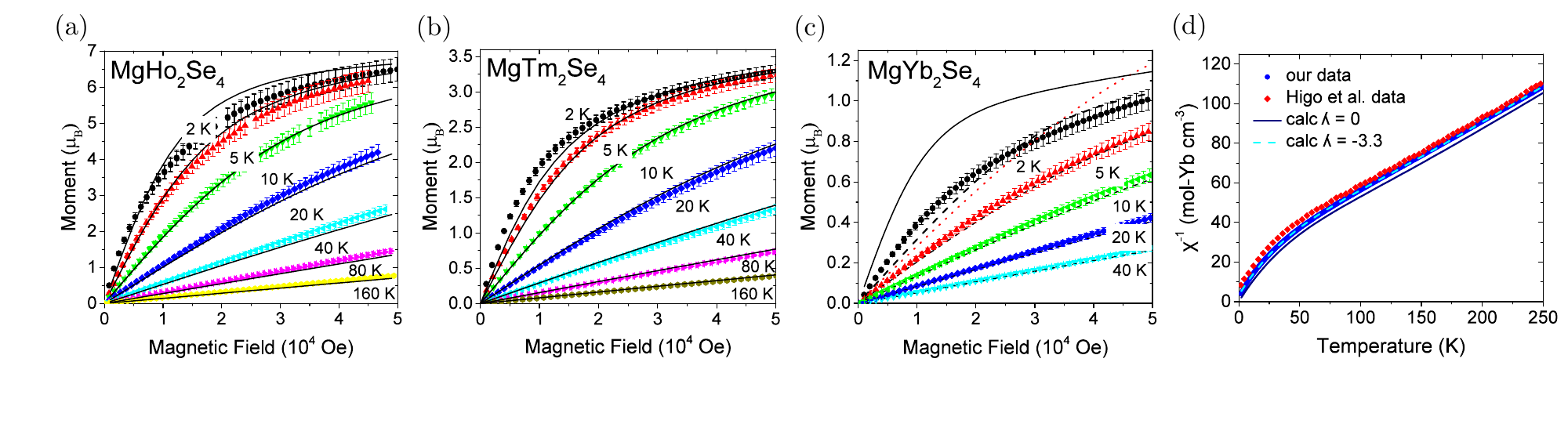}
\caption{Magnetization of MgRE$_2$Se$_4$ powders for RE = Ho, Tm, and Yb in panels (a), (b), and (c), respectively. Data taken at temperatures $T$ = 2, 5, 10, 20, 40, 80 and 160 K are shown as markers, error bars show systematic errors, calculated magnetization is shown as solid lines and dashed lines for non-interacting and interacting models, respectively.
The red dotted line in panel (c) is calculated from the CEF potential found in Ref.~\onlinecite{17_Higo_AYb2S4}.
Panel (d) shows inverse susceptibility of MgYb$_2$Se$_4$ in panel (d), with our data (H = 100 Oe) shown as circles, data from \onlinecite{17_Higo_AYb2S4} as diamonds, and the calculated susceptibility shown as a solid and dashed line for the noninteracting and interacting model, respectively. }
\label{fig:MVH}
\end{figure*}

\section{Magnetization}

To check the CEF potential found by refinement of the INS data, we performed a series of magnetization measurements as a function of both temperature and applied field, with main results shown in Fig.~\ref{fig:MVH}.
Symbols in this figure represent data, which is corrected for the demagnetizing field by assuming the powder sample is an oblate spheroid with a filling fraction of 60\%.
Solid lines in panels (a) -- (c) represent predictions of a non-interacting model using the Hamiltonian in Eq.~\ref{eq:CEF_D3d} with parameters B$_{nm}$ from Table~\ref{tab:BNM_energies} and an additional Zeeman term to account for the role of applied field. Solutions of this modified Hamiltonian were found by direct diagonalization with the field pointing along $x$, $y$ and $z$ directions. For each direction, the expected moment is calculated using Boltzmann statistics before averaging to simulate a powder.
This comprehensive approach is deemed more reliable than any that restricts attention to the ground state doublets only or treats the Zeeman term in the Hamiltonian perturbatively, as applied fields are known to both mix and shift the energy of low-lying excited CEF levels.

For all compounds, the measured and calculated magnetization show excellent agreement at high temperatures, as expected for strongly paramagnetic moments.
This agreement extends to all temperatures for MgTm$_2$Se$_4$, which has a ground state composed of momentless singlets. For MgHo$_2$Se$_4$ and MgYb$_2$Se$_4$ however, the calculated curve begins to overestimate the measured values at the lowest temperatures.
We attribute this discrepancy in MgYb$_2$Se$_4$ to the existence of net antiferromagnetic interactions, which are not accounted for in our independent moment CEF Hamiltonian. This conjecture is generally consistent with reports of negative Weiss constants in the literature on MgYb$_2$Se$_4$,  which range from $\Theta_{CW} = -9.2$~K\cite{17_Higo_AYb2S4} to  $\Theta_{CW} = -44$~K\cite{88_pawlak_Mag_CEF_Yb}, and reports of $\Theta_{CW} = -3.6(5)$~K\cite{15_Yaouanc_CdHo2S4} and $\Theta_{CW} = -7.6(2)$~K\cite{05_Lau_RE_chalcogenide_spinels} for CdHo$_2$Se$_4$, which is isostructural to MgHo$_2$Se$_4$. Though we caution against overinterpreting the results of Curie-Weiss fits in materials containing low-lying CEF transitions, these reports are sufficient to conclude antiferromagnetic interactions with an energy scale of a few Kelvin. For MgHo$_2$Se$_4$, we further note that the first two excited CEF levels (0.59 meV and 0.95 meV) are low enough in energy that interactions may mix these transitions with the ground state doublet and more fundamentally modify the effective spin state.

As a first step in exploring the role of interactions in these compounds, we also performed a series of self-consistent calculations using a Weiss molecular field, $\lambda$, and compared results to magnetization data for MgYb$_2$Se$_4$. Specifically, for a series of temperatures and applied fields, magnetization was defined as the solution to the transcendental equation $M = M_0(H + \lambda M)$, where $M_0$ is the calculated moment in the non-interacting model and $\lambda$ was determined by fitting to the data. In our analysis, we found $\lambda$=-3.4~mol-Yb~cm$^{-3}$ for MgYb$_2$Se$_4$. The curves associated with this analysis are shown as dashed curves in Fig.~\ref{fig:MVH}(c) and (d), and reveal a much improved match to both field and temperature data over the independent spin model. The current model is also greatly improved over calculations using CEF parameters of Ref.~\onlinecite{17_Higo_AYb2S4}, which concluded Heisenberg-like moments from fits of susceptibility vs temperature data. In particular, one can see that the more isotropic model, shown as a dotted red line for $T=2$~K, seems to be heading towards a much higher saturation moment than either the data or the predictions of the current paper.

The impact of interactions is further observed in the inverse susceptibility vs temperature curve, shown in Fig.~\ref{fig:MVH}(d) for MgYb$_2$Se$_4$. Here, we plot the calculated inverse susceptibility both without and with the interactions as solid and dashed curves respectively, along with data shown as blue circles. Whereas the non-interacting model prediction is systematically low, the curve including interactions matches the data quite well. In the same figure we also show the data from Ref.~\onlinecite{17_Higo_AYb2S4} as red diamonds, demonstrating consistency between the two data sets aside from a small offset which can attributed to a small amount of disorder. This punctuates the fact that both isotropic and highly anisotropic models are capable of describing magnetization versus temperature data at low fields, and higher field and spectroscopic measurements are absolutely essential if one wishes to determine information about the local CEF environment of local moments.

\section{Discussion and conclusions}

The current manuscript outlines the determination with INS of the symmetry-allowed CEF parameters for three members of the RE-spinel selenide family MgRE$_2$Se$_4$ (RE = Ho, Tm, and Yb). The parameters obtained are substantially different and demonstrably more accurate than previous efforts to determine CEF schemes by fitting magnetic susceptibility curves at low applied fields. This can be seen in the inability of parameters determined by the latter methods to either reproduce higher-field magnetization data or to successfully predict the energies of excited CEF transitions, which can be measured directly with INS\cite{88_pawlak_Mag_CEF_Yb,17_Higo_AYb2S4}.
In contrast, the parameters listed in Table~\ref{tab:BNM_energies} have been shown to largely reproduce the neutron scattering intensity as a function of both energy and temperature and magnetization data over a wide range of applied fields and temperatures. We can use these parameters to not only determine the ground state wavefunction of each material, as presented above, but also to revisit the role that ground state and exited levels have on low temperature magnetic properties.

For example, our measurements of MgHo$_2$Se$_4$ reveal a ground state Ising doublet with $m = 1.36(23)\mu_B$ moments and antiferromagnetic interactions, which may make this material amenable to a long-ranged ordered state similar to the one determines for CdHo$_2$Se$_4$\cite{15_Yaouanc_CdHo2S4}. Significantly however, we also observe several low lying excitations, including a singlet at 0.591(36) meV, a doublet at 0.945(30) meV, and a second singlet at 2.88(7) meV. This situation is reminiscent of the materials Tb$_2$Ti$_2$O$_7$ and Tb$_2$Sn$_2$O$_7$, where virtual transitions associated with low-lying CEF levels are strongly suspected to renormalize the effective Hamiltonian\cite{07_Molavian_virtual_CEF_excitations} and induce quantum fluctuations\cite{17_Takatsu_quadrupolar_states_Tb2Ti2O7,Takatsu_2011}. In the spinel selenides, the larger lattice parameters and rare earth to anion distances result in excited CEF energies even closer to the elastic channel, which implies an even faster timescale for quantum fluctuations.

In MgTm$_2$Se$_4$, the first excited CEF level contributes to the physics in a different way. Whereas our INS analysis concludes a simple singlet ground state, consistent with expectations\cite{80_Ben-Dor_CdRE2S_series,77_pokrzywnicki_mag_tm2se3_CdTm2Se4}, our INS spectra also reveals the existence of a low lying singlet at $E = 0.876(16)$~meV.  The Brillouin-function-like field dependence of magnetization in Fig.~\ref{fig:MVH}(b) demonstrates that the appreciable  occupation of this excited level by increasing either field or temperature endows the Tm$^{3+}$ atoms with a considerable finite moment, raising the intriguing possibility of stabilizing ordering phenomena at finite temperature with applied field, even as the system strives toward a singlet ground state at $T = 0$~K. Further, recent  theoretical work\cite{2019Changle_TFIM} has pointed out that the lowest two CEF levels form a quasi-doublet, allowing the magnetism in this material to be described by a transverse-field Ising model with a potential for an exotic quantum spin ice phase.

Only in MgYb$_2$Se$_4$ is the effective spin system well-isolated from the lowest excited level, at $E = 26.01(56)$~meV. A major insight of this work however is how highly anisotropic the Yb$^{3+}$ moments are in this system, which we infer not just from the analysis of our INS data but also from the saturation magnetization, which falls far short of expectations for isotropic spins.
The idea of strongly anisotropic Yb$^{3+}$ effective spins stands in contrast to earlier predictions of isotropic moments from magnetization\cite{17_Higo_AYb2S4} or weaker anisotropy from scaling arguments\cite{18_rau_Ba3Yb2Zn5O11}. The Yb pyrochlore-lattice materials stand out as rare examples where anisotropic interactions have been calculated and semiclassical phase diagrams have been produced as a function of material properties\cite{18_rau_Ba3Yb2Zn5O11}.
Thus, follow-up neutron diffraction measurements of low temperature ordered phases in this material might provide an opportunity to immediately test the validity of our results, and perhaps contribute to the understanding of the larger family of Yb$_2$M$_2$O$_7$ pyrochlores\cite{rau2019frustrated}.

Discussion of these three materials may naturally be grouped with consideration of sulfur (MgRE$_2$S$_4$) and cadmium (CdRE$_2$X$_4$) analogues\cite{65_flahaut,66_flahaut_erratum} and recent reports of classical spin-ice behavior in MgEr$_2$Se$_4$\cite{reig2018spin} and CdEr$_2$Se$_4$\cite{10_Lago_CdEr2Se4_heat_cap,18_Gao_CdEr2Se4_fennel_paper}. Together these publications show growing interest in the chalcogenide spinels, as a relatively unexplored family of highly frustrated magnets with a diversity of exotic states that rivals that of the 227 pyrochlore oxides. Proper consideration of local CEF environments is the first necessary step in modeling and fully understanding the associated physics.

\section{Acknowledgments}
\begin{acknowledgments}
This work was sponsored by the National Science Foundation, under grant number DMR-1455264-CAR.
D.R. further acknowledges the partial support of by the U.S. D.O.E., Office of Science, Office of Workforce Development for Teachers and Scientists, Office of Science Graduate Student Research (SCGSR) program.
The SCGSR program is administered by the Oak Ridge Institute for Science and Education for the DOE under contract number DE-AC05-06OR23100.
R. D. M. acknowledges the support of the Research Experience for Undergraduates program funded by the NSF under Grant No. 1659598.
Synthesis and magnetization measurements were carried out in the Materials Research Laboratory Central Research Facilities at the University of Illinois.
X-ray scattering measurements were performed at the Center for Nanophase Materials Sciences at Oak Ridge National Lab.
A portion of this work used resources at the Spallation Neutron Source, which is a DOE Office of Science User Facility operated by Oak Ridge National Laboratory.
\end{acknowledgments}

\bibliographystyle{apsrev4-1}

%

\end{document}